# Strain-stress study of Al$_x$Ga$_{1-x}$N/AlN heterostructures on *c*-plane sapphire and related optical properties

Yining. Feng[1,2], Vishal. Saravade[3], Ting-Fung. Chung[1], Yongqi. Dong[4,5], Hua. Zhou[5], Bahadir. Kucukgok[1,2], Ian. T. Ferguson[3,*], Na. Lu[1,6,*]

[1]Lyles School of Civil Engineering, Birck Nanotechnology Center, Purdue University, West Lafayette, IN 47907, USA

[2]Applied Materials Division, Argonne National Laboratory, 9700 South Cass Ave. Lemont, IL, 60439, USA

[3]Electrical and Computer Engineering, Missouri University of Science and Technology, Rolla MO 65409, USA

[4]National Synchrotron Radiation Laboratory, University of Science and Technology of China, Hefei, Anhui 230026, China

[5]Advanced Photon Source, Argonne National Laboratory, 9700 South Cass Ave. Lemont, IL, 60439, USA

[6]School of Materials Engineering, Purdue University, West Lafayette, IN 47907, USA

[*]corresponding authors: ianf@mst.edu ; luna@purdue.edu

This work presents a systematic study of stress and strain of Al$_x$Ga$_{1-x}$N/AlN with composition ranging from GaN to AlN, grown on a *c*-plane sapphire by metal-organic chemical vapor deposition, using synchrotron radiation high-resolution X-ray diffraction and reciprocal space mapping. The *c*-plane of the Al$_x$Ga$_{1-x}$N epitaxial layers exhibits compressive strain, while the *a*-plane exhibits tensile strain. The biaxial stress and strain are found to increase with increasing Al composition, although the lattice mismatch between the Al$_x$Ga$_{1-x}$N and the buffer layer AlN gets smaller. A reduction in the lateral coherence lengths and an increase in the edge and screw dislocations are seen as the Al$_x$Ga$_{1-x}$N composition is varied from GaN to AlN, exhibiting a clear dependence of the crystal properties of Al$_x$Ga$_{1-x}$N on the Al content. The bandgap of the epitaxial layers is slightly lower than predicted value due to a larger tensile strain effect on the *a*-axis compared to the compressive strain on the *c*-axis. Raman characteristics of the Al$_x$Ga$_{1-x}$N samples exhibit a shift in the phonon peaks with the Al composition. The effect of strain is also discussed on the optical phonon energies of the epitaxial layers. The techniques discussed here can be used to study other similar materials.

**Introduction**

  III-Nitride alloys have attracted considerable attention in a wide range of applications of optical, optoelectronic, high-power, and high-frequency devices such as light emitting diodes (LEDs), laser diodes, and high electron mobility transistors (HEMTs)[1–6]. For instance, the hexagonal $Al_xGa_{1-x}N$ is one of the most promising candidates for ultraviolet (UV)-LED applications, especially because of its wide bandgap ($E_g$) range from 3.42 eV (for GaN) to 6.2 eV (for AlN) at room temperature[7]. $Al_xGa_{1-x}N$ is also an optimum intermediate layer for InGaN-based LEDs and InAlN transistors[2,8–10]. $Al_xGa_{1-x}N$/GaN HEMTs paves the way for achieving high power radio frequency (RF) devices due to high electron mobility, large critical breakdown field, high sheet charge density, high electron saturation velocity, and high temperature operation[11]. $Al_xGa_{1-x}N$ /AlN heterostructure combines the photodetector abilities of deep ultra-violet (DUV) AlN along with the tunable bandgap $Al_xGa_{1-x}N$, thereby also suppressing the visible spectrum and enhancing the UV/visible rejection ratio[12,13]. This has applications in military target or missile detection, biochemical sensing, as solar-blind detectors, air/water purification, curing, and biomedical therapies and instrumentation[14–17]. Considering the photodetector applications, AlN has a higher bandgap than $Al_xGa_{1-x}N$ and hence the resulting photodetector spectrum (of light waves incident on $Al_xGa_{1-x}N$ surface) would be dominantly dependent on the $Al_xGa_{1-x}N$ epitaxial layer properties; as opposed to other structures consisting of $Al_xGa_{1-x}N$ and a lower bandgap material, where it could be difficult to separate the effects of the two materials on the energy spectrum. Also, an $Al_xGa_{1-x}N$ /AlN structure would have the flexibility to function as a photodetector from top and bottom sides with front and back illumination respectively, with the top $Al_xGa_{1-x}N$ epitaxial layer having bandgap range from ~4 eV to ~6 eV depending on the Al content, and a bottom AlN layer with 6.2 eV band gap. Using an AlN intermediate layer for $Al_xGa_{1-x}N$ could also improve the crystal quality of the heterostructure and reduce absorption losses[16]. However, highly efficient and reliable electronic and optoelectronic devices require epitaxial layers with excellent crystal quality (i.e., low dislocation density and residual strain). It is challenging to grow high-quality $Al_xGa_{1-x}N$ thin films, particularly with high Al composition ($x$); this is due to the lattice mismatch and thermal expansion difference between the thin films and substrates, which generally results in high-level strain-stress and mosaicity[18–20]. Strain-stress in epitaxial layers is one of the leading factors that reduces the electron mobility and degrades the device performance[21–23]. Also, their optical and morphological properties could be improved by reducing the strain and stress. Therefore, it is vital to understand the strain and stress mechanism for improving the optical and electronic properties and applications of III-Nitrides.

  High-resolution X-ray diffraction (HRXRD) and reciprocal space mapping (RSM) could be used to understand the crystal properties and to analyze the strain and stress in epitaxially grown III-Nitride films[24]. The effect of different intermediate layers such as AlN, GaN, and step-graded $Al_xGa_{1-x}N$ for $Al_xGa_{1-x}N$ /GaN HEMT structures on silicon (111) substrate has been studied by XRD, RSM and Hall effect measurements, showing that the in-plane stress can largely affect the two-dimensional electron gas mobility and carrier concentration[25]. The origin of stresses in $Al_xGa_{1-x}N$/GaN heterostructures grown on *c*-plane sapphire substrate relies mainly on the thickness and growth temperature of the layers, alloy composition, and doping. However, a systematic study of strain and stress in $Al_xGa_{1-x}N$/AlN heterostructures, especially for high $x$ (>0.5) Al$_x$Ga$_{1-x}$N epitaxial layers, on $c$-plane sapphire substrates by synchrotron radiation HRXRD and RSM technique has not been reported. It is crucial to study the crystal properties of Al$_x$Ga$_{1-x}$N /AlN structures, which is a step towards improving their quality and potential for practical applications.

In this work, the overall strain, biaxial strain, hydrostatic strain, and biaxial stress along the $a$- and $c$-axis, are analyzed and calculated for Al$_x$Ga$_{1-x}$N/AlN heterostructure on sapphire substrates with varying $x$ and AlGaN composition from GaN to AlN using synchrotron radiation HRXRD and RSM. The effect of the Al content on the crystal properties, dislocation densities and coherence lengths are discussed. The effect of strain on the optical properties of the Al$_x$Ga$_{1-x}$N thin films has been investigated using photoluminescence (PL) and Raman spectroscopy.

**Results and Discussion**

The crystal structure and lattice parameters of MOCVD-grown Al$_x$Ga$_{1-x}$N and AlN have been studied using HRXRD and RSM techniques, while photoluminescence and Raman measurement results are discussed to understand the bandgap and phonon modes in Al$_x$Ga$_{1-x}$N and AlN. Fig. 1 shows the 2θ-ω Bragg reflections ($\lambda$=1.23984 Å) around (0002) crystal planes for Al$_x$Ga$_{1-x}$N with varying $x$. The effect of strain is taken into account to determine the $x$ values as per the synchrotron radiation HRXRD results[26]. Bragg reflection peaks of (0002) from Al$_x$Ga$_{1-x}$N and AlN, and of (0006) from the sapphire substrate, are observed.

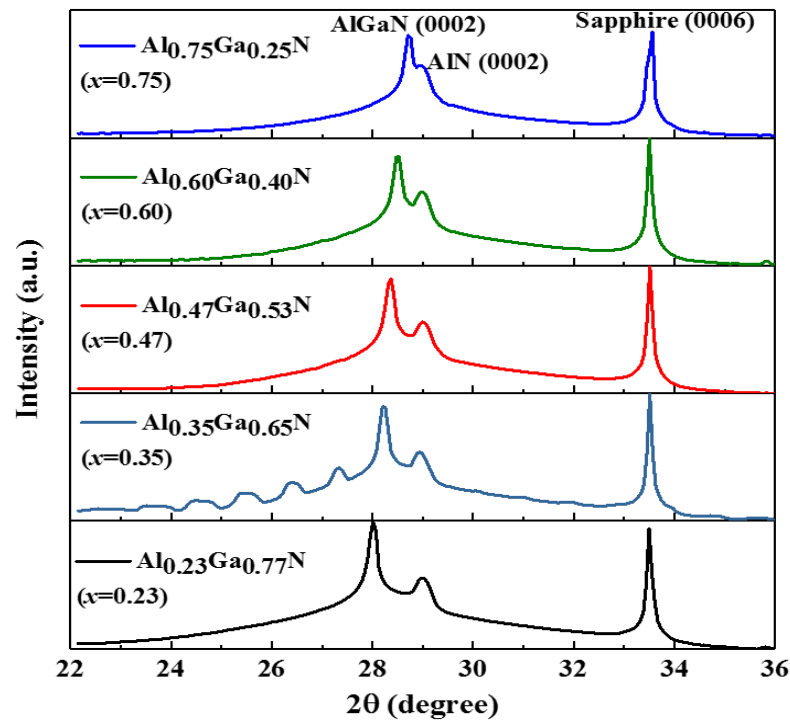

**Figure 1**. HRXRD 2θ-ω scan near (0002) Bragg reflection plane for the Al$_x$Ga$_{1-x}$N thin films

The diffraction peak position of Al$_x$Ga$_{1-x}$N (0002) shifts towards higher 2θ with an increase in $x$, which implies smaller c-axis lattice constant with an increasing Al content in Al$_x$Ga$_{1-x}$N. At high $x$, the Al$_x$Ga$_{1-x}$N peak starts to overlap with the AlN (002) peak. The c-axis lattice parameter seems to reduce from GaN to AlN as the Al content in Al$_x$Ga$_{1-x}$N increases.

A FWHM of about ~180 arcsec and a vertical coherence length of ~100 nm was observed for the $Al_xGa_{1-x}N$ layers, where the FWHM seems to increase and the vertical coherence length to reduce with $x$. For the AlN peak, a vertical coherence length of ~36 nm with a FWHM of ~360 arcsec was seen. The AlN (0002) peak from the intermediate layer is not affected with the Al content, as the Al incorporation takes place in the top $Al_xGa_{1-x}N$ layer. In the rest of the paper, samples with $x = 0.23, 0.47, 0.75$ that cover a range of $x$ to reasonably understand the structural and optical properties of $Al_xGa_{1-x}N$ /AlN are discussed.

The out-of-plane $c$-axis lattice constant ($c$) of $Al_xGa_{1-x}N$ thin films were calculated as shown in Table I. Vegard's law provides reliable unstrained lattice constants ($c_0, a_0$) for $Al_xGa_{1-x}N$ films using the bandgaps of GaN and AlN, and considering the very small lattice mismatch (~2%) between GaN and AlN[19,26–29]. The calculated $c$, is lower than the unstrained $c_0$, indicating a compressive strain along the $c$-axis (out-of-plane) in the $Al_xGa_{1-x}N$ thin films.

RSM based analysis were also done to determine the lattice constants and the stress-strain phenomenon in $Al_xGa_{1-x}N$ with changes in $x$. Fig. 2 shows the symmetric plane RSM in the (0002) direction. A clear broadening of $Al_xGa_{1-x}N$ reciprocal lattice points (RLPs) reflection intensity distribution towards $Q_z$ and $Q_x$ is seen. It can be observed that the maximum reflection intensity of $Al_xGa_{1-x}N$ shifts to higher $Q_z$ values and the lattice constant $c$ reduces, as $x$ increases, which agrees very well with the results obtained from the 2θ-ω scan. Also, broadening along the $Q_z$ direction increases with $x$. Changes in the RSM plots with different Al content seem to be dominated by the $Al_xGa_{1-x}N$ layer.

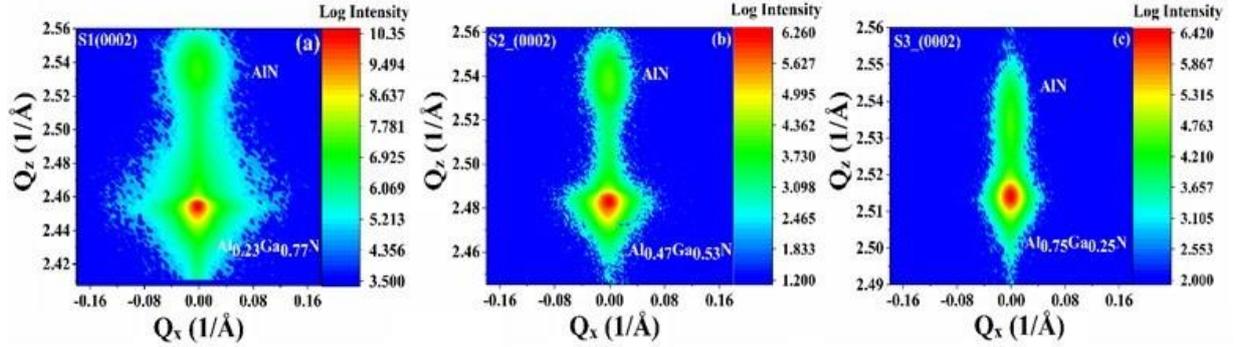

**Figure 2**. Symmetric RSM (0002) scan of the $Al_xGa_{1-x}N$ thin film.

Reciprocal space map around the AlN asymmetric ($10\bar{1}3$) RLP is illustrated in Fig. 3. Based on the information from the asymmetric RSM scan, lattice parameters ($a$ and $c$) were calculated for the hexagonal structure Eq. (1)[30–32]:

$$a = \frac{2\pi}{|Q_x|}\sqrt{\frac{4(h^2+k^2+hk)}{3}}, c = \frac{2\pi l}{Q_z}, \qquad (1)$$

Table I presents the calculated lattice parameters from the asymmetric RSM measurement (in this particular case, $h=1$, $k=0$, and $l=3$) for $Al_xGa_{1-x}N$. The calculated $c$ from asymmetric RSMs is very close to the one obtained by HRXRD 2θ-ω scans for each sample, with a difference of about 0.06%; hence only the c-parameters from the HRXRD results are shown. The calculated $a$ is larger than the unstrained one ($a_0$) obtained by Vegard's law, which is due to the tensile strain along the $a$-axis (in-plane) in the $Al_xGa_{1-x}N$ epitaxial layers. Also, the $a$-lattice constant reduces with an increase in $x$, similar to $c$. A reduction in the lattice size and increase in the strain is seen in Al$_x$Ga$_{1-x}$N with an increase in the Al content in the alloy.

Fig. 3 shows that with increasing Al composition, the maximum reflection intensity of Al$_x$Ga$_{1-x}$N RLPs progressively shifts from a partially relaxed ($R=1$) towards a fully strained ($R=0$) position. Since the AlN layer is thinner (~120 nm) than the Al$_x$Ga$_{1-x}$N layer (~800 nm), its reflection peak intensity is lower than Al$_x$Ga$_{1-x}$N. The intensity of Al$_x$Ga$_{1-x}$N RLP broadens along the direction associated with the relaxation of the layer (the dashed black line). The Al$_x$Ga$_{1-x}$N RLPs get closer to the fully strained position with an increase in $x$ as seen in Fig. 3. Note that both AlN and Al$_{0.75}$Ga$_{0.25}$N have a similar $Q_x$ value of -2.38 Å$^{-1}$. An increase in the strain is observed with Al incorporation in Al$_x$Ga$_{1-x}$N, despite of reductions in lattice mismatch. The broadening in the symmetric and asymmetric RLPs implies an increase in the screw and edge dislocations (which are in the order of $10^8$-$10^9$ cm$^{-2}$) respectively with $x$. The RSM and the 2θ-ω results show that the dislocations and the coherence lengths in Al$_x$Ga$_{1-x}$N/AlN change with $x$. Lattice constants of hexagonal AlN are typically smaller than GaN and hence, a reduction in the lateral correlation lengths and an increase in the dislocations are seen as the Al$_x$Ga$_{1-x}$N composition is varied from GaN to AlN.

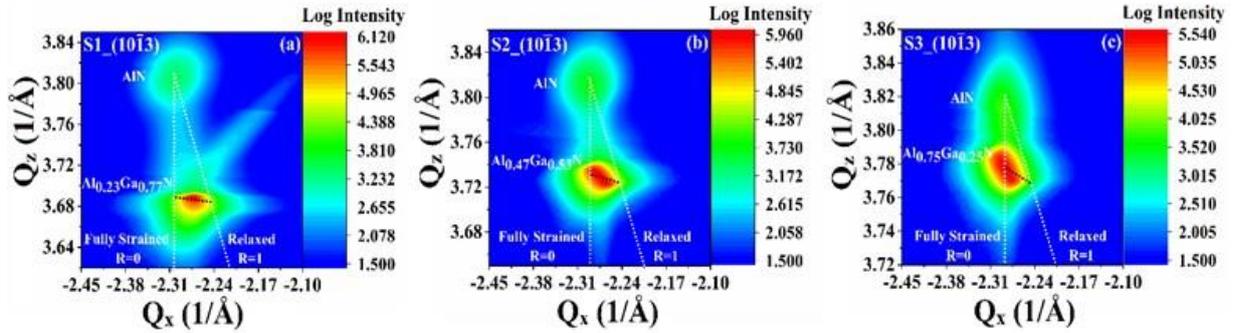

**Figure 3.** Asymmetric RSM (10$\bar{1}$3) scan of the Al$_x$Ga$_{1-x}$N thin films grown on sapphire. (a) Al$_{0.23}$Ga$_{0.77}$N, (b) Al$_{0.47}$Ga$_{0.53}$N, and (c) Al$_{0.75}$Ga$_{0.25}$N. The dashed white lines indicate where the fully relaxed ($R=1$) and fully strained ($R=0$) Al$_x$Ga$_{1-x}$N layers with varying Al compositions should be. The dashed black lines show the relaxation directions in the reciprocal space for different Al compositions.

The overall in-plane strain ($\varepsilon_a$) and out-of-plane strain ($\varepsilon_c$) in the Al$_x$Ga$_{1-x}$N layers were determined using Eq. (2)[31,33–35]:

$$\varepsilon_a = \frac{a-a_0}{a_0}, \quad \varepsilon_c = \frac{c-c_0}{c_0}, \tag{2}$$

The calculated strains ($\varepsilon_a$ and $\varepsilon_c$) are attributed to the biaxial ($\varepsilon_a^b$ and $\varepsilon_c^b$) and hydrostatic ($\varepsilon_h$) strains as shown in Eq. (3)[34,36]. ($\varepsilon_a^b$ and $\varepsilon_c^b$ are the biaxial strains along $a$- and $c$-directions, respectively.)

$$\varepsilon_a = \varepsilon_a^b + \varepsilon_h, \quad \varepsilon_c = \varepsilon_c^b + \varepsilon_h, \tag{3}$$

where $\varepsilon_h$ is defined as $\varepsilon_h = \frac{1-v}{1+v}(\varepsilon_c + \frac{2v}{1-v})$, $v$ is Poisson's ratio of Al$_x$Ga$_{1-x}$N calculated using Vegard's law ($v_{AlGaN}(x) = x\,v_{AlN} + (1-x)\,v_{GaN}$)[37] and shown in Table I. For the hexagonal crystal structure, the in-plane biaxial stress ($\sigma^b$) in the Al$_x$Ga$_{1-x}$N epitaxial layer can be determined by $\sigma^b = M_b \varepsilon_a^b$, where $M_b$ is the biaxial elastic modulus given by $M_b = (C_{11} + C_{12} + 2\frac{C_{13}^2}{C_{33}})$[34]. The elastic constants ($C_{ij}$) of Al$_x$Ga$_{1-x}$N (Table I) can be obtained by Vegard's law ( $C_{ij}^{AlGaN}(x) = xC_{ij}^{AlN} + (1-x)C_{ij}^{GaN}$ ) .[38,39] The calculated strains, biaxial strains, hydrostatic strain, and biaxial stress for $Al_xGa_{1-x}N$ epitaxial layers are summarized in Table II. It can be seen that the in-plane (biaxial) strains are tensile, while the out-of-plane (biaxial) strains are compressive because of the different lattice mismatch along the in-plane and out-of-plane axes[19] as also seen in the HRXRD results.

The biaxial strain has values close to the total strain in AlGaN due to the relatively smaller values of $\varepsilon_h$ and very few impurities introduced during growth. Also, the full width at half maximum (FWHM) values of the HRXRD (0002) ω scans (not shown here) are found to be 627, 642, and 847 arcsec for $Al_{0.23}Ga_{0.77}N$, $Al_{0.47}Ga_{0.53}N$, and $Al_{0.75}Ga_{0.25}N$, respectively (Table III)[32]. The lateral coherence lengths would range from 100 nm to 200 nm and have inverse proportionality with the Al content, indicating that the $Al_xGa_{1-x}N$ samples used in this study are of good crystal quality.

The broadening of the FWHM of (0002) HRXRD ω scans in AlGaN could be associated with the screw (*c*-type) threading dislocation (TD) along the *c*-axis. Fig. 4(a) presents the compositional dependence of screw (*c*-type) TD density and out-of-plane strain in the $Al_xGa_{1-x}N$ thin films. The dislocation density of the $Al_xGa_{1-x}N$ thin films can be estimated from[40]:

$$D_{screw} = \frac{\beta_{(0002)}^2}{4.35 b_{screw}^2}, \quad (4)$$

where $D_{screw}$ is the screw type TD, $\beta$ is the FWHM of the (0002) ω scan, and $b_{screw}$=5.1855 Å is the Burgers vector length for screw-type TD. As *x* increases, both the screw type TD density and the strain increase (Fig. 4(a)). Evidently, the high density of screw dislocation observed in the Al-rich samples originated from a compressive strain along the *c*-axis (up to 0.6%) and a biaxial stress (up to 6.313 GPa), in $Al_xGa_{1-x}N$, as presented in Table III.

Photoluminescence measurements (Fig. 4(b)) further indicate and help to understand the strain and stress in the epitaxial layers. A broadening of the $Al_xGa_{1-x}N$ peaks in observed with an increase in *x*. Also, there is a shift in the peak positions compared to the unstrained energy gaps that are predicted by Vegard's law. The PL peak positions are measured at 3.88, 4.27, and 5.25 eV for $Al_{0.23}Ga_{0.77}N$, $Al_{0.47}Ga_{0.53}N$, and $Al_{0.75}Ga_{0.25}N$, respectively. According to Vegard's law, the predicted energy gap values for *x*=0.23, 0.47, and 0.75 are 4.06, 4.73, and 5.51 eV respectively (considering $E_g(AlN)$=6.2 eV, $E_g(GaN)$=3.42 eV). Smaller bandgap in the measured samples as compared to the predicted values, could be attributed more to the stronger tensile strain effect along the *a*-axis direction than the *c*-axis compressive strain ($\varepsilon_a \approx 2\varepsilon_c$) in the $Al_xGa_{1-x}N$ epitaxial layers and hence, to the overall larger lattice constants of AlGaN epitaxial layers as compared to unstrained $Al_xGa_{1-x}N$. Also, the bandgap increases with *x* as would be expected and seems to be tunable between GaN and AlN. The PL peak broadening, intensity suppression and peak shifts could have multiple origins such as a statistical variation in the composition, Al-induced alloy disorder, strain and dislocations.

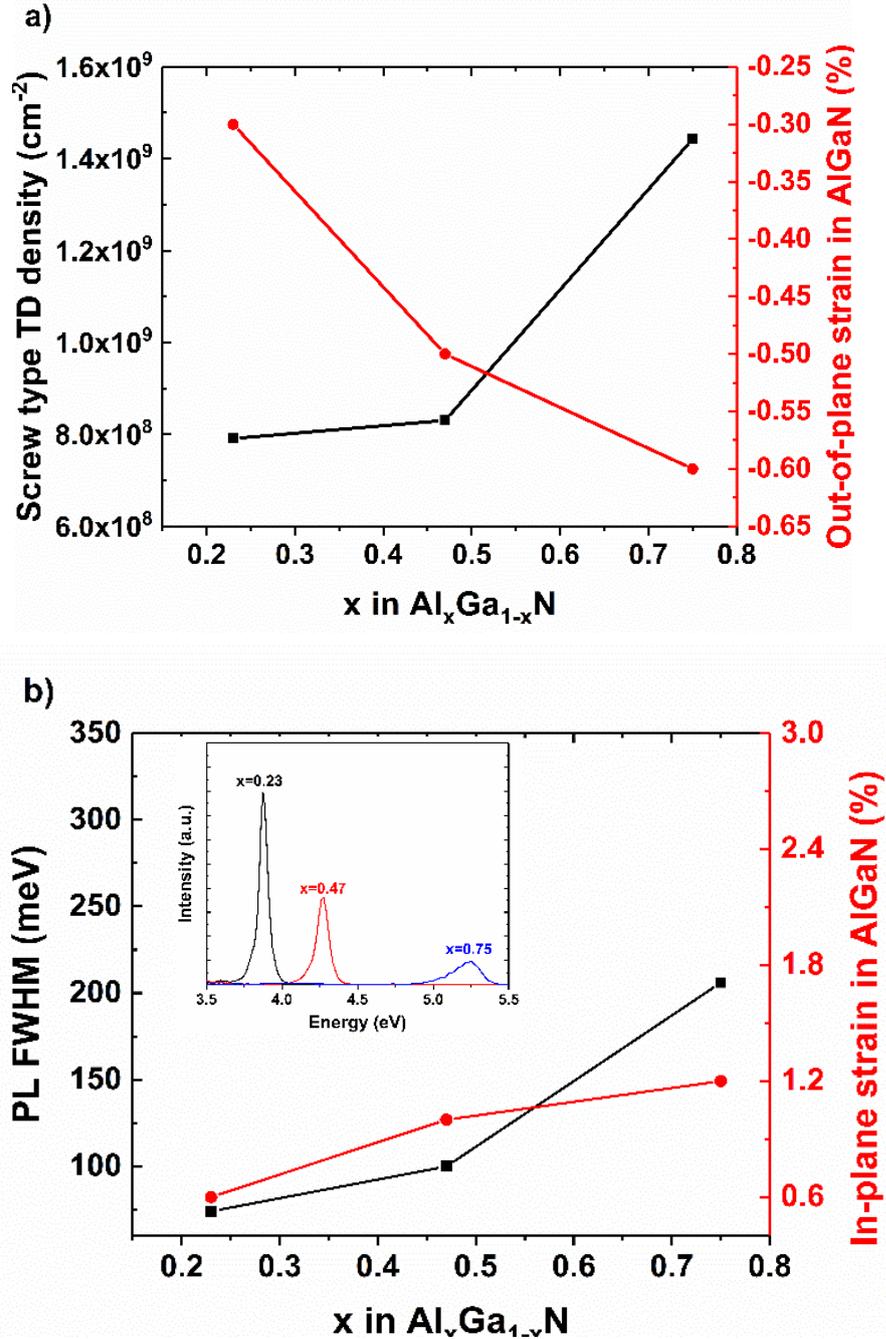

**Figure 4**. Compositional dependence of (a) screw (*c*-type) TD density and out-of-plane strain, (b) PL FWHM and in-plane strain of Al$_x$Ga$_{1-x}$N layers. The inset shows the room temperature PL spectra of Al$_{0.23}$Ga$_{0.77}$N, Al$_{0.47}$Ga$_{0.53}$N, and Al$_{0.75}$Ga$_{0.25}$N

Raman spectra of the Al$_x$Ga$_{1-x}$N samples under 532 nm excitation are shown in Fig. 5. Two-mode behavior for the E$_2^{high}$ phonon[37] and one-mode behavior for the A$_1^{LO}$ phonon[41] are seen. Here, E$_2^{high}$ and A$_1^{LO}$ phonon modes correspond to the atomic oscillations in the *c*-plane (parallel to the *a*-axis) and along the *c*-axis, respectively. The phonon peaks exhibit a shift with increasing *x*. The E$_2^{high}$ (GaN-like) phonon is located at 575, 587, and 607 cm$^{-1}$ for *x*=0.23, 0.47, and 0.75, respectively, while the E$_2^{high}$ (AlN-like) phonon is located at ~650 cm$^{-1}$ with a weak composition dependence. The A$_1^{LO}$ phonon also exhibits strong composition dependence, from 783 to 864 cm$^{-1}$ when $x$ increases from 0.23 to 0.75. A sharp peak at 750 cm$^{-1}$ (marked with an asterisk) and a weak peak at 576 cm$^{-1}$ (marked with an asterisk and most visible for $x$=0.75 because the peak is overlaid by the strong $E_2^{high}$ (GaN-like) peak) correspond to phonon vibrations of the the sapphire substrate. The composition-dependence behavior of the $E_2^{high}$ (GaN-like) and $A_1^{LO}$ modes is in good agreement with previous work on Al$_x$Ga$_{1-x}$N epitaxial layers[42–44] wherein the Raman results also confirm the wurtzite structure of the Al$_x$Ga$_{1-x}$N layer with its hexagonal [0001] crystal plane parallel to the $c$-plane sapphire substrate. Strain due to alloying seems to be the major mechanism for the observed Raman shifts (the difference in phonon energies due to substrate-induced strain is small). Moreover, the $E_2^{high}$ (AlN-like) peak intensity varies with $x$, as the phonon vibrations are sensitive to atom compositions. Therefore, higher $x$ values revealed more distinct $E_2^{high}$ (AlN-like) phonon vibration peaks, which is typical of alloy semiconductors. The result also suggests that the AlN buffer layer quality is good, so there is a small substrate-induced strain in the Al$_x$Ga$_{1-x}$N epitaxial layers.

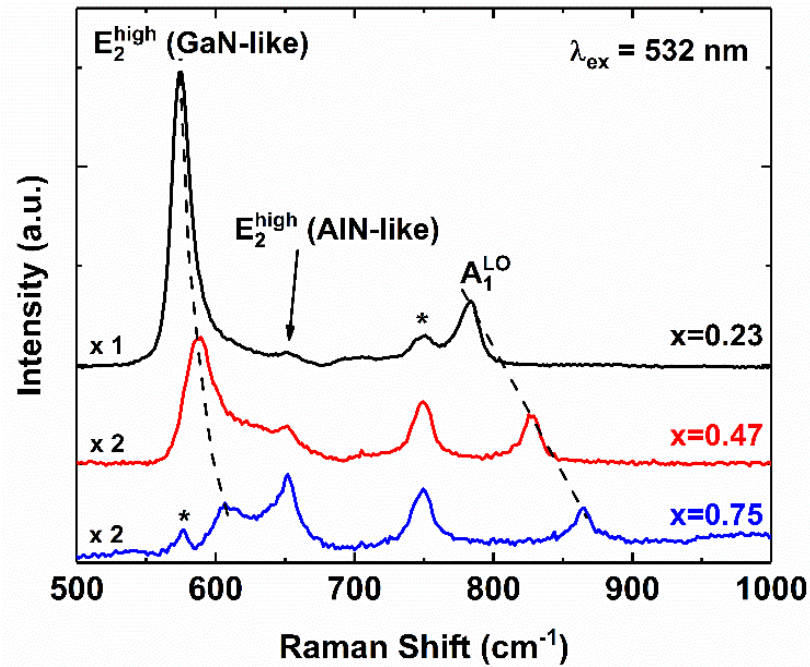

**Figure 5.** Raman spectra for Al$_x$Ga$_{1-x}$N thin films ($x$=0.23, 0.47, 0.75) measured with a 532 nm excitation laser under ambient conditions. The Raman spectra for $x$=0.47 and 0.75 are multiplied by a factor of two for clarity. The dashed lines marking the composition dependence of the $E_2^{high}$ (GaN-like) and $A_1^{LO}$ modes are guides to the eye. Asterisks near 576 (only observable for $x$=0.75 because of overlapping with the $E_2^{high}$ (GaN-like) mode) and 750 cm$^{-1}$ show the $c$-plane sapphire substrate phonons.

**Conclusion**

In summary, the study focuses on the strain-stress status of Al$_x$Ga$_{1-x}$N epitaxial layer grown by MOCVD on a $c$-plane sapphire substrate with AlN as an intermediate layer. The lattice parameters reduce as the Al content in Al$_x$Ga$_{1-x}$N is increased. The out-of-plane strain of Al$_x$Ga$_{1-x}$N is compressive, and the in-plane strain is tensile. The strain increases with $x$, even though the lattice mismatch between Al$_x$Ga$_{1-x}$N and AlN reduces. Broadening of the RSM peaks and the HRXRD rocking curve scans imply a consistent reduction in correlation lengths and higher dislocation densities with increasing $x$ as the $Al_xGa_{1-x}N$ composition is varied from GaN to AlN. The bandgap of $Al_xGa_{1-x}N$ increases with $x$, as expected. Also, the values are smaller than the unstrained bandgap predicted by Vegard's law, due to a larger tensile strain on the $a$-axis compared to the compressive strain on the $c$-axis. The $E_2^{high}$ and LO phonons exhibit a shift with an increasing $x$ caused due to the strain accompanied with alloying. Considering the potential of $Al_xGa_{1-x}N$ for optical and electronic applications, this work adds towards the understanding of crystal and optical properties of $Al_xGa_{1-x}N$/AlN structure with high $x$; which need to be addressed or utilized for the development of optimum $Al_xGa_{1-x}N$/AlN based devices. Also, the techniques discussed here could be applied towards studying other materials wherein the crystal defects seem to be a hindrance towards utilizing the material.

## Methods

**Metal-organic chemical vapor deposition (MOCVD) Growth.** $Al_xGa_{1-x}N$ thin films with varying $x$ were grown on $c$-plane sapphire substrates by metal-organic chemical vapor deposition (MOCVD). The precursors for Al, Ga, and N, are trimethylaluminum (TMA), trimethylgallium (TMG), and ammonia ($NH_3$), respectively. To remove surface contamination, sapphire substrates were heated at 1100 °C in $H_2$ ambient prior to the growth. A 40 Torr chamber pressure was maintained for the growth of AlN and $Al_xGa_{1-x}N$ epitaxial layers. A ~20 nm low-temperature (LT) AlN nucleation layer with a V/III ratio of 3000 was deposited on the sapphire substrate at 600 °C. The temperature was then increased to 1040 °C to grow a ~100 nm high-temperature (HT) AlN buffer layer. Finally, a ~800 nm $Al_xGa_{1-x}N$ epitaxial layer was grown on the AlN layer at 1140 °C.[3] The samples were cooled in $NH_3$ environment.

**Materials Characterizations**. Synchrotron radiation HRXRD measurement were performed at 33IDD beamline at the Advanced Photon Source, Argonne National Laboratory. It is equipped with a standard six-circle Kappa-type diffractometer and Pilatus 100 K area detector. A deep ultraviolet (DUV) PL spectroscopy (excitation at 224 nm) was used to measure the optical properties of the $Al_xGa_{1-x}N$ thin films. Micro-Raman spectroscopy was performed using a Horiba Jobin-Yvon Xplora confocal Raman spectrometer in a backscattering configuration with a 532 nm excitation laser and a grating of 1800 lines/mm.

## Data Availability

The datasets generated during and/or analyzed in the current study are available from the corresponding author on reasonable request.

**Acknowledgements**

The authors at Purdue University are grateful for the financial supports from National Science Foundation CAREER program (under Grants of CMMI – 1560834) and NSF IIP- 1700628. The authors thank Professor Yong P. Chen at Purdue University for allowing the use of Raman spectroscopy equipment in his lab. This research used resources of the Advanced Photon Source, a U.S. Department of Energy (DOE) Office of Science User Facility operated for the DOE Office of Science by Argonne National Laboratory under Contract No. DE-AC02-06CH11357.

**Author Contributions**

N. L. and I. F. contributed to the conception and design of experiments. Y. F., V. S., T. C., B. K., conducted most of experiments, characterization and drafted the manuscript. Y. D and H. Z. conducted synchrotron and reciprocal space mapping. N. L, I. F. and H. Z. edited and revised the manuscript. N. L. and I. F. supervised the project. All the authors discussed the results.

**Additional Information**

**Competing interests:** The authors declare no competing interests.

**Tables**

**TABLE I.** Calculated strained ($a$, $c$) parameters (from HRXRD 2θ-ω scan and asymmetric RSM scans) and unstrained lattice parameters ($a_0$, $c_0$) (from Vegard's law), Al composition ($x$)[26], elastic constants ($C_{11}$, $C_{12}$, $C_{13}$, and $C_{33}$) and Poisson ratio ($v$) of Al$_x$Ga$_{1-x}$N. ($v_0^{AlN}$=0.207 and $v_0^{GaN}$=0.202)

| Al composition ($x$) | In-plane lattice parameter [Å] | | Out-of-plane lattice parameter [Å] | | Elastic constant [GPa] | | | | Poisson ratio ($v$) |
|---|---|---|---|---|---|---|---|---|---|
| | Calculated ($a$) | Unstrained ($a_0$) | Calculated ($c$) | Unstrained ($c_0$) | $C_{11}$ | $C_{12}$ | $C_{13}$ | $C_{33}$ | |
| $x$=0.23 | 3.190 | 3.171 | 5.121 | 5.138 | 394.83 | 145.92 | 104.39 | 395.93 | 0.203 |
| $x$=0.47 | 3.185 | 3.152 | 5.061 | 5.088 | 399.87 | 146.88 | 102.71 | 393.77 | 0.204 |
| $x$=0.75 | 3.169 | 3.130 | 4.998 | 5.031 | 405.75 | 148.00 | 100.75 | 391.25 | 0.205 |

**TABLE II.** Measured in-plane and out-of-plane strains, biaxial strains, hydrostatic strain, and biaxial stress of Al$_x$Ga$_{1-x}$N. Positive and negative values denote tensile and compressive strains respectively

| Al composition ($x$) | In-plane strain ($\varepsilon_a$) [%] | In-plane biaxial strain ($\varepsilon_a^b$) [%] | Out-of-plane strain ($\varepsilon_c$) [%] | Out-of-plane biaxial strain ($\varepsilon_c^b$) [%] | Hydrostatic strain ($\varepsilon_h$) | Biaxial stress ($\sigma^b$) [GPa] |
|---|---|---|---|---|---|---|
| $x$=0.23 | 0.6 | 0.6 | -0.3 | -0.3 | 1.06×10$^{-6}$ | 2.9 |
| $x$=0.47 | 1.0 | 1.0 | -0.5 | -0.5 | -2.35×10$^{-5}$ | 5.1 |
| $x$=0.75 | 1.2 | 1.2 | -0.6 | -0.6 | -3.50×10$^{-6}$ | 6.3 |

**TABLE III.** Summary of structural and optical results of the Al$_x$Ga$_{1-x}$N thin films

| Al composition ($x$) | FWHM of HRXRD [arcsec] | Screw TD Density [cm$^{-2}$] | FWHM of PL [meV] | Energy gap [eV] |
|---|---|---|---|---|
| $x$=0.23 | 627 | 7.91×10$^8$ | 74 | 3.88 |
| $x$=0.47 | 642 | 8.30×10$^8$ | 100 | 4.27 |
| $x$=0.75 | 847 | 1.44×10$^9$ | 206 | 5.25 |